\documentclass[journal,letterpaper,twocolumn]{IEEEtran}
\usepackage{cite}
\usepackage{fancyhdr}
\usepackage{float}
\usepackage{graphicx}
\usepackage{epsfig}
\usepackage{bm}
\usepackage{amsmath}
\usepackage{mathrsfs}
\usepackage{subfigure}
\usepackage{array}
\usepackage{booktabs}
\usepackage{amssymb}
\usepackage{graphicx}
\usepackage{color}
\usepackage{epstopdf}
\usepackage{algorithm}
\usepackage{amsthm}
\usepackage{bbm}
\usepackage{graphicx}
\usepackage{epsfig}

\usepackage{cite}
\makeatother

\usepackage{algpseudocode}
\usepackage{amsmath}
\title{When Federated Learning Meets Blockchain: A New Distributed Learning Paradigm}

\begin{document}
\author{\IEEEauthorblockN{Chuan Ma, \emph{Member, IEEE},  %\thanks{This work is supported in part by the National Key R$\&$D Program under Grants 2018YFB1004800, and by National Natural Science Foundation of China under 61872184 and 61727802. The corresponding authors are Jun Li and Feng Shu.}
\thanks{C. Ma, J. Li and L. Shi are with the School of Electronic and Optical Engineering, Nanjing University of Science and Technology, Nanjing, China (e-mail: \{chuan.ma, jun.li\}@njust.edu.cn and slong1007@gmail.com).}
Jun Li, \emph{Senior Member, IEEE},
%\thanks{J. Li is also with the Department of Software Engineering, Institute of Cybernetics, National Research Tomsk Polytechnic University, Tomsk, 634050, Russia.}
Ming Ding, \emph{Senior Member, IEEE},\\
\thanks{M. Ding is with Data61, CSIRO, 2015, Australia (e-mail: Ming.Ding@data61.csiro.au).}
%Howard H. Yang, \emph{Member, IEEE},
%\thanks{H. H. Yang and T. Q. S. Quek are with the Information System Technology and Design Pillar, Singapore University of Technology and Design (e-mail: \{howard yang, tonyquek\}@sutd.edu.sg).}
%Tony Q. S. Quek, \emph{Fellow, IEEE},
Long Shi, \emph{Member, IEEE},
Taotao Wang, \emph{Member, IEEE},\\ \thanks {T. Wang is with the College of Electronics and Information Engineering, Shenzhen University, Shenzhen 518060, China (e-mail: ttwang@szu.edu.cn).}
Zhu Han, \emph{Fellow, IEEE}, \thanks{Z. Han is with the Department of Electrical and Computer Engineering, University of Houston, Houston, TX 77004, USA (e-mail: zhan2@uh.edu)}
and H. Vincent Poor, \emph{Fellow, IEEE}
\thanks{H. V. Poor is with the Department of Electrical Engineering, Princeton University, Princeton, NJ 08544, USA (e-mail: poor@princeton.edu).}
}}

\maketitle
\begin{abstract}
Motivated by the explosive computing capabilities at end user equipments, as well as the growing privacy concerns over sharing sensitive raw data, a new machine learning paradigm, named federated learning (FL) has emerged. By training models locally at each client and aggregating learning models at a central server, FL has the capability to avoid sharing data directly, thereby reducing privacy leakage. However, the traditional FL framework heavily relies on a single central server and may fall apart if such a server behaves maliciously. To address this single point of failure issue, this work investigates a blockchain assisted decentralized FL (BLADE-FL) framework, which can well prevent the malicious clients from poisoning the learning process, and further provides a self-motivated and reliable learning environment for clients. In detail, the model aggregation process is fully decentralized and the tasks of training for FL and mining for blockchain are integrated into each participant. In addition, we investigate the unique issues in this framework and provide analytical and experimental results to shed light on possible solutions.
\end{abstract}
\section{Introduction}
Future wireless networks are featured by low latency and high reliability. Thus, machine learning (ML) embedded in each device is a ravishing solution that each user equipment (UE) has the capability to make decisions by its local data, even when it loses connectivity to the wireless system. Since the data at each device is limited, the training of on-device ML models always requires the data exchange among UEs \cite{9048613}.

However, directly exchanging data among UEs may cause serious risks in privacy leakage and information hijacking \cite{8274963}. To reduce this risk, federated learning (FL) is proposed, which is a new ML framework that trains an AI model across multiple UEs holding local datasets. In details, FL allows to train machine learning models locally at distributed UEs; after that, the UEs share the parameters of the locally trained models to a central server (i.e., the aggregator) where a global model is aggregated. Therefore, the UEs under the FL framework have the capability to cooperatively learn a global model without exchanging their data directly. Moreover, FL has been applied to real-world applications, including health care and autonomous driving \cite{9076082}. %A typical FL framework can be found in Fig.~1, which consists of a central sever and several local clients.

Although FL shows its effectiveness in preserving privacy, it still endures several limitations. First, in the FL process, the single centralized aggregator is assumed to be trustworthy and it shall make fair decisions in terms of the user selection and aggregation. However, this assumption is not always appropriate, especially in the real-world operations. This is because a biased aggregator can intentionally emerge prejudice to a few selected UEs, thereby damaging the learning performance \cite{9048613}. Second, the aim of FL is restricted to applications orchestrated by the centralized aggregator. As a result, the resiliency of an aggregator depends on the robustness of the central server, and a failure in the aggregator could collapse the entire FL network. Then, although local data is not explicitly shared in the original format, it is still possible for adversaries to reconstruct the raw data approximately, especially in the aggregation process. In particular, privacy leakage may happen during model aggregating by outsider attacks. Lastly, the existing design is vulnerable to the malicious clients that might upload poisonous models to attack the FL network \cite{9084352}.

As a secure technology, blockchain has the capability to tolerate single point failure with distributed consensus, and it can further implement incentive mechanisms to encourage participants to effectively contribute to the system \cite{8733825}. Therefore, blockchain is introduced to FL to solve its limitations mentioned above. In \cite{8733825}, a blockchained FL architecture was developed to verify the uploaded parameters and it investigated the related system performances, such as the learning delay and the block generation rate. Moreover, work \cite{8843900} proposed a privacy-aware architecture that uses blockchain to enhance security when sharing parameters of machine learning models with other UEs. In addition, the authors in \cite{8905038} proposed a high-level but complicated framework by enabling encryption during model transmission and providing incentives from participants, and the work \cite{SHARMA2020102220} further applied this framework in the defensive military network. With the advanced features of blockchain such as tamper-proof, anonymity and traceability, an immutable audit trail of ML models can be created for greater trustworthiness in tracking and proving provenance \cite{9051184}. In addition, security and privacy issues of the decentralized FL framework are investigated in \cite{9134967,inproceedings1,8843900}, which delegate the responsibility of storing ML models to a trust community in the blockchain. %{\color{blue}add a table to compare the existing BC-FL works with ours and highlight the key differences}
However, the assumption on the trust community may infer the same privacy issue when ML models transmitting over air, and the credibility of this community also needs further verification. In addition, these works have either not clearly clarified and fully addressed the incident issues, such as the long learning delay and the impact of blockchain forking on FL, or have difficulty in application.
Thus, in this work we have fully detailed the whole process of blockchain assisted decentralized FL (BLADE-FL), which has the capability to overcome the single point of failure problem. In addition, we further investigate the residual issues that exist in the BLADE-FL framework, and provide related solutions.
%introduce a new FL framework called blockchain-aided FL (BC-FL) that uses blockchain to replace the central server.
%The main benefits brought from blockchain are that it provides confidential updates from local training nodes and feeds back corresponding rewards to accelerate cooperation.
In detail, we present the design of the BLADE-FL framework in Sec.~II, and residual issues including privacy, resource allocation and lazy clients are investigated in Sec.~III. In Sec.~IV we provide extensive experimental results to show the effectiveness of the corresponding solutions. Finally, future directions and conclusion are drawn in Sec.~V.

\section{The framework of BC-FL}
With the aid of blockchain, we aim to build up a secure and reliable FL framework. To ensure this, the model updating process of FL is decentralized at each participating client, which is robust against the malfunction of traditional aggregators.
In this article, we detail the BLADE-FL framework, to achieve a dynamic client selection and a decentralized learning aggregation process.

The BLADE-FL framework is composed of three layers. In the network layer, the network features a decentralized P2P network that consists of task publishers and training clients, wherein a learning mission is first published by a task publisher, and then completed by the cooperation of several training clients. Different from previous work that model aggregation happens in a trust community in the blockchain \cite{8733825,8843900,8905038,SHARMA2020102220,9051184,9134967,inproceedings1}, we realize a fully decentralized framework that each client needs to train ML models and mine blocks for publishing aggregating results. In the blockchain layer, each FL-related event, such as publishing a task, broadcasting learning models, and aggregating learning results, is tracked by blockchain. In the application layer, the SC and FL are utilized to execute the FL-related events. Next, we will detail the working flow and key components of the BLADE-FL framework.
%The main components are dynamic client negotiation, FL modules in Blocked-FL, blockchain operation in Blocked-FL and Blocked-FL operation, and the details will be illustrated in the following subsections.
%\begin{figure*}
%\centering
%  \includegraphics[width=0.65\textwidth]{framework.pdf}
%  \caption{The framework of the proposed BC-FL system} \label{framework}
%\end{figure*}

\subsection{Working Flow}
\begin{figure*}
\centering
  \includegraphics[width=0.77\textwidth]{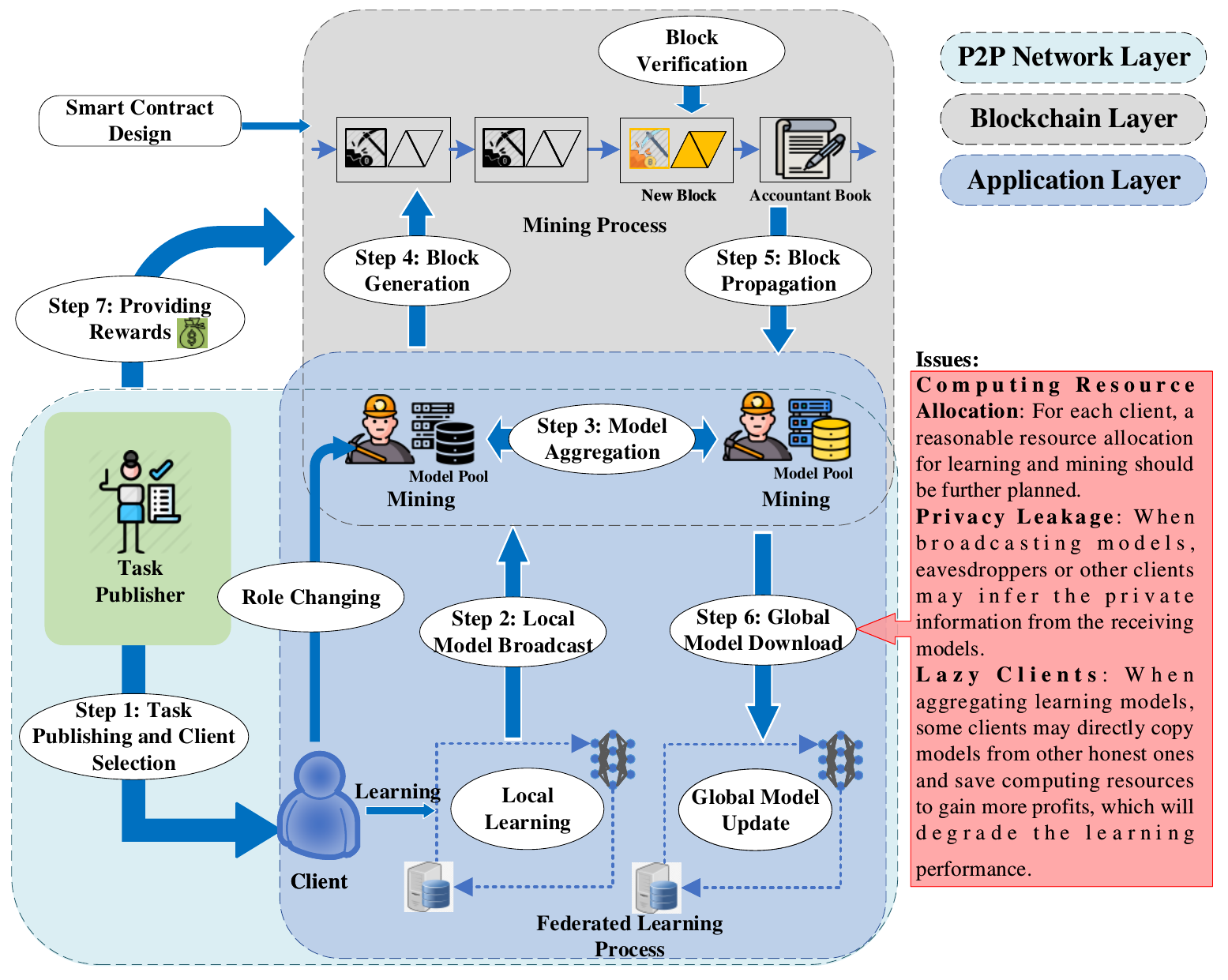}
  \caption{The working flow of the blockchain assisted decentralized federated learning (BLADE-FL)} \label{system}
\end{figure*}

As shown in Fig.~\ref{system}, the working flow of the proposed framework operates in the following steps:
%The working flow of the proposed framework operates in the following steps\footnote{Mentioned here, more details will be given in \textbf{Sec. III-C}.}:
\begin{itemize}
  \item \textbf{Step 1}: Task publishing and node selection. A task publisher broadcasts a FL task through deploying a SC over the blockchain network. In the deployed SC, the task publisher needs to deposit reward as financial  incentives to the learning task. The SC selects available training nodes to participate in this learning task.
  \item \textbf{Step 2}: Local model broadcast. Each training client runs its local training by using its own data samples and broadcasts its local updates and the corresponding processing information (e.g., computation time and local data size) over the P2P network. Privacy leakage may happen during this transmission, and we further investigate this issue in Sec.~III-A.
  %\item Step 3: Local model upload. Training nodes broadcast the local updates and corresponding information (e.g., computation time and local data size) over the P2P network. For more details, please refer to \textbf{Sec. III-C}.
      %and associate with an available miner in the network. %In this case that clients are the miners as well, this step can be skipped.
  \item \textbf{Step 3}: Model aggregation. Upon receiving the local updates from other training nodes before a preset time-stamp, each client updates the global model according to the aggregating rule defined in the SC.
  \item \textbf{Step 4}: Block generation. Each training client changes roles from trainer to miner and begins mining until either it finds the required nonce or it receives a generated block from other miners. The learning results are stored in the block as well. When one miner generates a new block, other clients verify the contents of this block (e.g., the nonce, the state changed by SC, the transactions, and the aggregated model). The resource allocation issue happens in each client in this step, and related discussions will be given in Sec.~III-B.
  \item \textbf{Step 5}: Block propagation. If a block is verified by the majority of clients, this block will be added on the blockchain and accepted by the whole network. The lazy client issue happens in this step and we further investigate it in Sec.~III-C.
  \item \textbf{Step 6}: Global model download and update. Each training client downloads the aggregated model from the block and performs updates before the next round of learning.
  \item \textbf{Step 7}: Reward allocation. The SC deployed by the task publisher rewards the training clients according to their contributions in the learning task.
\end{itemize}

Before delving into each step, we elaborate on key designs in the BLADE-FL as follows.

\subsection{Smart Contract Design}
Smart contracts are self-executing contracts defining rules for negotiating, verifying the fulfilment of rules and executing the agreement using the formal code.
The BLADE-FL framework relies on SC to enable trusted dynamic client selections in terms of desired distributed learning services, without relying on a centralized authority.
Moreover, BC-FL enables all clients to verify the learning results that are recorded on the blockchain, whereby distributed clients can be incentivized to participate and untrusted learning models can be detected. Based on the verification results, the reputation of each distributed client can be automatically updated, making the selection of learning nodes more reliable.
In addition, the design of SC in the BC-FL also includes the aggregating rules, and thus provides a fair and open rewarding feedback for participating clients. The SC in BC-FL enables three main functions as follows:

\textbf{Function 1}: Learning task publishing. A task publisher broadcasts a FL task through SC to all users. The SC contains the task requirements (e.g., the data size, training accuracy, latency, etc.), the aggregating rules and rewards paid by the task publisher.

\textbf{Function 2}: Dynamic bidding for requests and automatic incentive. Distributed training nodes, acting as auctioneers, bid for the task by replying their costs and capabilities. Note that in order to enforce accountability, each training client has to stake a deposit to the SC. The task replies from training nodes are recorded on the blockchain by the SC. Then the SC selects training clients with  more valuable replies (e.g., higher capability and lower cost) as the bid winners to jointly execute the FL task. The training clients that lose the bidding will reclaim their deposits from the SC, while the deposits made by winners will be automatically refunded if the learning results are verified to be trustworthy afterward.

\textbf{Function 3}: Learning results aggregation and rewards feedback. Before generating a new block, each client will aggregate the uploaded models according to the aggregating rule in SC, in which the contribution of each one in the aggregated model is also recorded in the newly generated block. Then SC is automatically triggered to reward the miner that helps aggregate the learning model and the training clients that contribute to the FL process.
\subsection{The BLADE-FL Design}
The main purpose of the BLADE-FL is to enable a trusted cooperative machine learning among distributed nodes. The decentralized accountability enables all miners to verify the quality of uploaded models that are recorded on the blockchain. In addition, distributed training nodes can be motivated to participate in the FL process and misbehaving ones can be recognized from providing low quality of FL services. The key steps are illustrated as follows:

\textbf{Local model update and upload}: Training nodes are bid winners with capable devices and available sets of data samples. In each learning iteration, each training node updates a local ML model in a parallel manner by using the global model and its local data samples, and broadcasts its local model in the network. This article considers that local updates can be received by all miners through the gossip protocol \cite{jelasity2011gossip} over the P2P network. In this context, the aggregation process in the traditional FL is decentralized to each client that stores the uploaded models in its model pool, respectively.

\textbf{Model aggregation}: After collecting the uploaded models in the pool, each client calculates the global model updates according to the aggregating rule in SC. In the proposed architecture, the clients are designed to aggregate the learning parameters truthfully through a distributed ledger. Similar to the prevailing block structure in \cite{8843900}, each block in a ledger consists of the body and header parts. Concretely, the body stores the local model updates, such as the local data size and computing time of the associated training node and the aggregated learning parameters. The header contains the information of a pointer to the previous block, block generation rate, and the output value, such as the proof of work (PoW), in the consensus protocol.

\textbf{\textbf{Model recording and publishing}}: The clients record the the aggregated models in their block and publish the recorded models by broadcasting the generated block to the whole network. The blocks can be generated by using distributed or lightweight consensus protocols, such as PoW, proof of stake (PoS), delegated PoS (DPoS), etc. \cite{8869754}. In this article, we consider PoW due to its strong security over decentralized networks. This article uses a synchronous schedule to ensure that all miner start mining at the same time
%\footnote{Future work can be considered here that miners are running in an asynchronous way.}.
Once a client find the hash value, its candidate block becomes to be a new block, and this block generation rate is controlled by the PoW difficulty. Then, this generated block is broadcasted to all other miners in the framework. All the other miners need to verify the nounce and the aggregated results contained in this block. For example, clients can compare the aggregated results with the one in the publishing block or use a public testing dataset to justify the effectiveness of the uploaded models. If the verification result is correct, other clients will accept it as a legal block and record it; otherwise, others will discard this generated block and continue to mine on the previous legal block.

\textbf{Reward allocation}:
The task publisher provides learning rewards for the participating training nodes, and the volume can be proportional to the size of training data. It is noted that the reward mechanism can be further mended by combining consider the data size and the quality of data samples. In this case, clients are responsible to verify the trustworthiness of local updates after aggregation,
to address the situation that untruthful UEs may exaggerate their sample sizes with abnormal local model updates. Specifically, when clients calculate the rewards to each training node, they can give scores/reputations to the training nodes based on the model qualities. In the next aggregation, nodes with low scores will be given less weights, and identified and gradually ignored during the learning. In practice, this can be guaranteed by Intel's software guard extensions, allowing applications to be operated within a protected environment, which has already been used in the blockchain technologies \cite{8048837}.
In addition, miners can also obtain rewards from mining and aggregating models, which can be treated as a gas tax in the traditional blockchain.
\section{Unique Issues and potential solutions}
In this section, we describe three critical issues that the proposed framework may confront with, namely privacy, resource allocation, and lazy clients%, and their potential solutions in Table~\ref{table2}.

%\begin{table*}
%\newcommand{\tabincell}[2]{\begin{tabular}{@{}#1@{}}#2\end{tabular}}
%\centering
%\caption{A summary of issues and probable solutions in the proposed BC-FL framework}
%\begin{tabular}{|cl|l|}
%  \hline
%  {}&Issues&Probable Solutions\\
%  \hline
%  Multi-functional Miner &\tabincell{l} {$\bullet$ Computing Resource Allocation\\ $\bullet$ Lazy nodes}&
%  \tabincell{l}{$\bullet$ Optimization needs accurate mathematical models\\ $\bullet$ Incentive mechanisms can be designed to discourage/penalize lazy \\~~nodes} \\
%  \hline
%  {Cross Verification}&\tabincell{l} {$\bullet$ Learning Verification\\ $\bullet$ Mining Verification}&
%   \tabincell{l}{$\bullet$ Digital Signature System: recognize malicious clients by verifying \\~~learning results\\ $\bullet$ Reputation System: reduce verification delay of high-reputation miner}\\
%  \hline
%  Forking&\tabincell{l} {$\bullet$ Training nodes process local learning \\using different base learning models}&
%  \tabincell{l}{$\bullet$ The highest reputation rule: store model from the highest miner\\ $\bullet$ The smallest ED rule: use ED to abandon other branches}\\
%  \hline
%\end{tabular}
%\label{table2}
%\end{table*}
\subsection{Privacy}
In the BLADE-FL, the roles of each client includes mining and training. To aggregate the global model, the trained local model will be published among clients which raises privacy issues. Previous works \cite{8733825,8843900,8905038,SHARMA2020102220,9051184,9134967,inproceedings1}, usually artificially assign the training and mining tasks to two disjoint sets of clients, and widely adopt that the miners are always trustful. However, if there exists an eavesdropper in the wireless environment, the published information of local models can cause privacy leakage. To address this, a differentially private mechanism can be implemented at the client side. In detail, keys steps are listed as follows:
\begin{itemize}
  \item Each client sets up a self-required privacy level for itself before training. For example, the $i$-th client may have a local privacy budget $\epsilon_i$. Note that a small value of $\epsilon_i$ represents a high local privacy level, and will induce more additive noises on the parameters.
  \item To achieve a local differential privacy (LDP), each client will add a random noise which follows a certain distribution on the uploaded models. For example, a random Gaussian noise $N(0, \sigma^2)$ or a Laplace noise $Lap(\lambda)$ will be added. Note that, a large noise power, i.e., $sigma^2$ implies a high privacy level.
  \item Upon receiving the perturbed models, all clients can aggregate the global model locally, and store it in the generated block. Because of the injected noise, the learning convergence as well as the system performance will be negatively affected. A tradeoff between the privacy requirement and the learning performance needs further investigation. In addition, an non-uniform allocation of additive noise over communication rounds may improve the learning performance. For example, a decay rate for the noise power can be applied when the learning accuracy between two adjacent communication rounds stops improving \cite{9347706}.
\end{itemize}
\subsection{Computing Resource Allocation}
Since the computation resource is limited at each client, each participant needs to appropriately allocate the resources for local training and mining to complete the task. Specifically, more computing resources can be devoted to either faster model update or block generation. To meet the specific task requirements, such as learning difficulty, accuracy, and delay, each node optimizes its allocation strategy to maximize its reward under constraints of local capability.

According to the constraints, the computation resource allocation can be formulated as an optimization problem under the accurate mathematical model. In details:
\begin{itemize}
  \item The block generation rate is determined by the computation complexity of the hash function and the total computing power of the blockchain network (i.e., total CPU cycles). The average CPU cycles required to generate a block can be defined as $kc_{\textrm{B}}$, where $k$ denotes the mining difficulty, and $c_{\textrm{B}}$ denotes the average number of total CPU cycles to generate a block. Thus, the average generation time of a block ($t_{\textrm{B}}$) can be expressed as $\frac{kc_{\textrm{B}}}{Nf}$, where $N$ is the number of clients, and $f$ denotes the CPU cycles per second of each client.
  \item The training time consumed by each training iteration $t_{\textrm{T}}$ can be expressed as $\frac{|D|c_{\textrm{T}}}{f}$, where $|D|$ denotes the number of samples of each client, and $c_{\textrm{T}}$ denotes the number of CPU cycles required to train one sample.
  \item Considering that a typical FL learning task is required to be accomplished within a fixed duration of $T_{\textrm{Sum}}$, it should satisfy that $K (\tau t_{\textrm{T}}+t_{\textrm{B}})\leq T_{\textrm{Sum}}$, %the number of local training iterations $\tau$ can be expressed as $\tau=\frac{1}{r_{\textrm{T}}}(\frac{T_{\textrm{Sum}}}{K}-r_{\textrm{B}})$,
      where $K$ denotes the total communication round, and $\tau$ denotes the local training epoches. Thus, to achieve a required learning performance, an appropriate choice for the communication round $K$ should be investigated under a certain ratio between the computing and mining time.
\end{itemize}

\subsection{Lazy nodes}
As the verification is processed locally, a lazy client may not perform local learning and directly copy uploaded parameters from other clients to save its computing resource. As a result, the client can devote more mining resources to reaping more mining rewards with a higher probability. However, this action significantly degrades the network learning performance. To investigate the effect of lazy nodes on the system performance, we provide related experimental results in Sec.~V-D.

To address the lazy client issue, we can implement a signature process at each client, which is based on the pseudo-noise (PN) sequence. Note that the signature mechanism here is completely different from the digital signature. What we need is a signature that is resilient to noise perturbation because the lazy clients are likely to perturb the plagiarized local models to hide the misbehavior. This process will introduce a negligible burden to the system but can provide a high detection accuracy. In details,
\begin{itemize}
  \item Before broadcasting the local updates, each client will produce a PN sequence with a length $L$, where $L$ is usually a very large number (larger than the number of model parameters) and we select a same length with model parameters and add them to the updates. This PN sequence has a high self-correlation coefficient and is hard to detect or re-produce by other clients. At least, the complexity of detecting the PN sequence should be much larger than that of training the neural network so as to deter the attempt to discover the used PN sequence.
  \item Upon receiving local updates from the other clients, each client will use its own PN sequence to check the correlation coefficient with the updates. If there exists high peaks in terms of the cross-correlation coefficient, then the lazy clients will be detected.
  \item Once a lazy client is recognized by a local client, this client can publish the previously used PN sequence to others and invite other honest clients to verify this process. Then any future updates from the lazy client might be discarded as punishments.
\end{itemize}

\section{Experimental Results and Probable Solutions}
In this section, we provide some experimental results to show the issues in the multi-functional miner in the proposed BLADE-FL system.
\subsection{System setup}
For each experiment, we first divide the original training data into non i.i.d. training sets, locally compute a stochastic gradient descend (SGD) update on each dataset, and then aggregate updates to train a globally shared classifier. We evaluate the prototype on the Fashion-MNIST dataset and Cifar-10 data. %The provided dataset in MNIST is divided into 60,000 training examples and 10,000 testing examples.
In the following results, we collect 20 runs for each experiment and record the average results.
%The global epoch is set to 300 iterations at the server side, while 120 iterations are implemented at each client side, and the local batch size is set to 1200.
For the blockchain setup, we set the total computation resource $T_{\textrm{Sum}}=200$ for each training node, and the total number of clients is set to $N=20$. In each communication round, each client uses $t_\textrm{B}$ time resources to generate a block and $t_\textrm{T}$ time resources to pursue a learning epoch, where $t_B=2$ for all experiments. Let $\theta=t_\textrm{T}/t_\textrm{B}$, and a larger $\theta$ implies that the client spares more computing resources to learning in each communication round.

\subsection{Investigation on the local differential privacy}
\begin{figure}
\centering
  \includegraphics[width=0.45\textwidth]{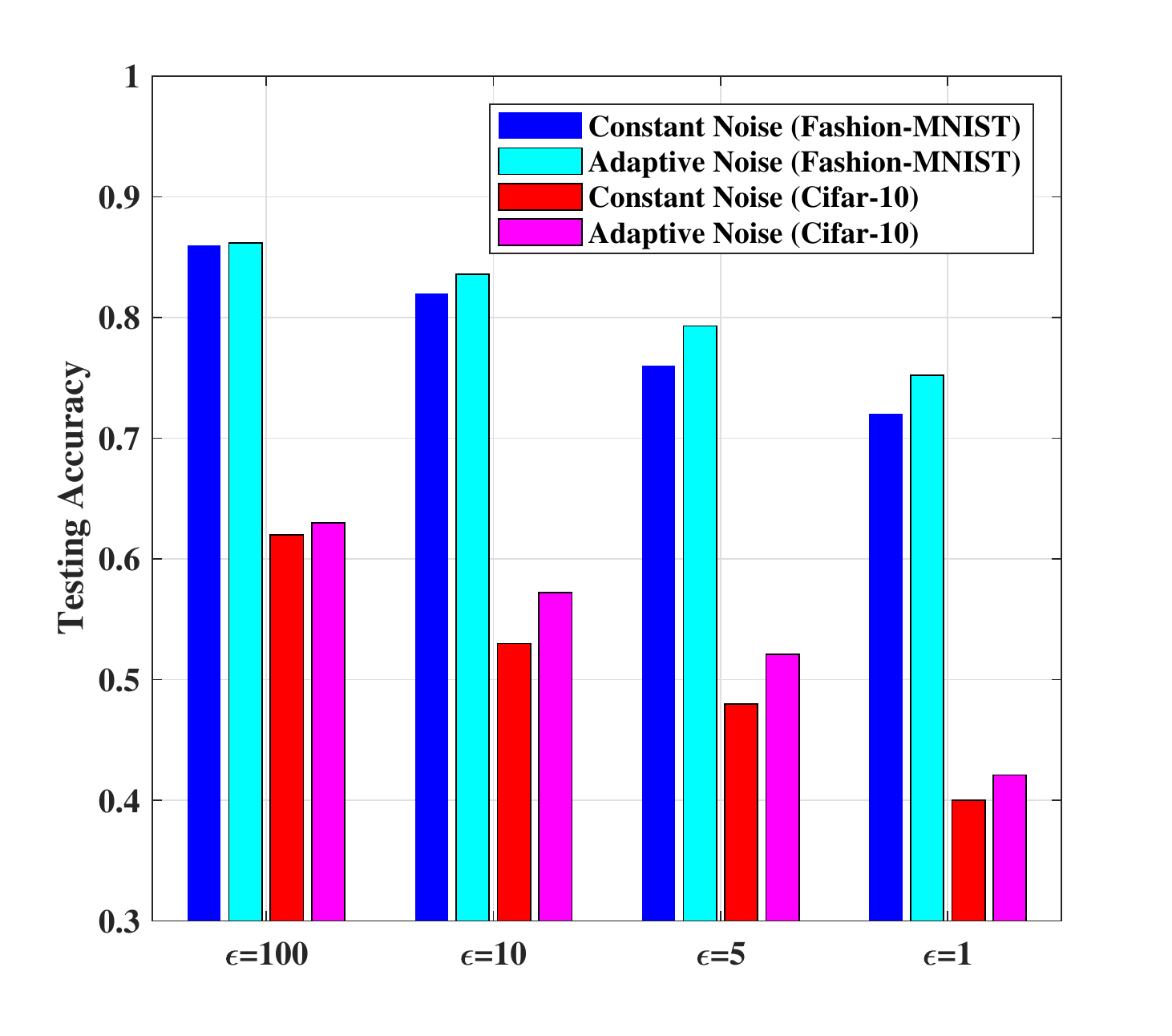}
  \caption{The learning performance with respect to different privacy levels} \label{p}
\end{figure}

In this subsection, we apply local differential privacy on each client by adding random Gaussian noises on the uploaded models in each communication round. The testing accuracies of the Fashion-MNIST and Cifar-10 dataset are plotted in Fig.~\ref{p} with respect to different privacy levels $\epsilon$. In addition, an adaptive noise decaying method is compared with the constant one, which will decrease the noise power when the accuracy stopes increasing.
As can be observed in this figure, the system achieves a higher performance with a larger value of $\epsilon$, which is under a weaker privacy protection, and the adaptive method can further improve the learning performance under the same level of privacy protection.
\subsection{Investigation on the resource allocation}
\begin{figure}
\centering
\subfigure[Fashion-Mnist]{\label{loss1}
  \includegraphics[width=0.23\textwidth]{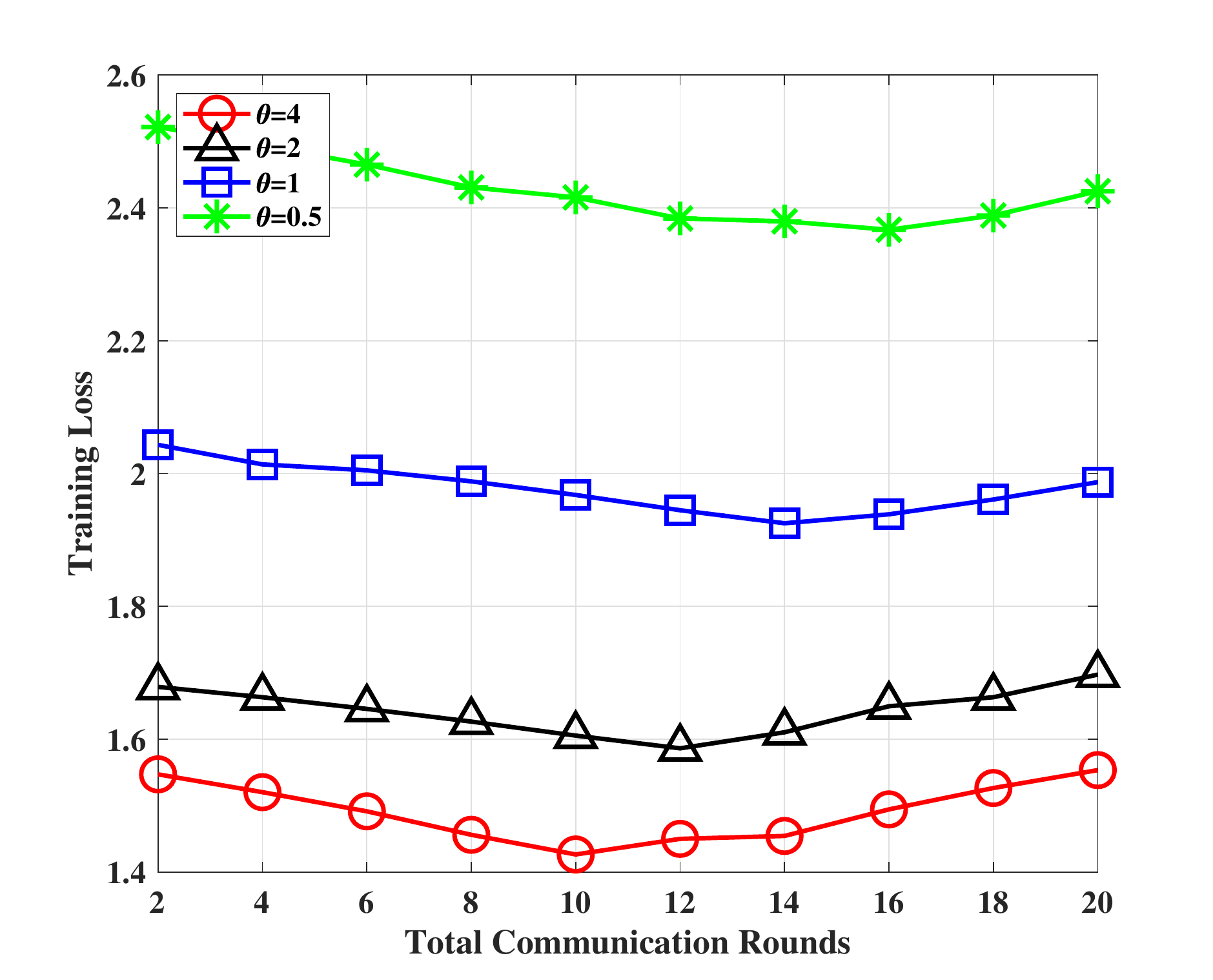}}
 % \caption{The loss function of different total communication rounds for different ratios} \label{rs1}
\subfigure[Cifar-10]{\label{loss2}
  \includegraphics[width=0.23\textwidth]{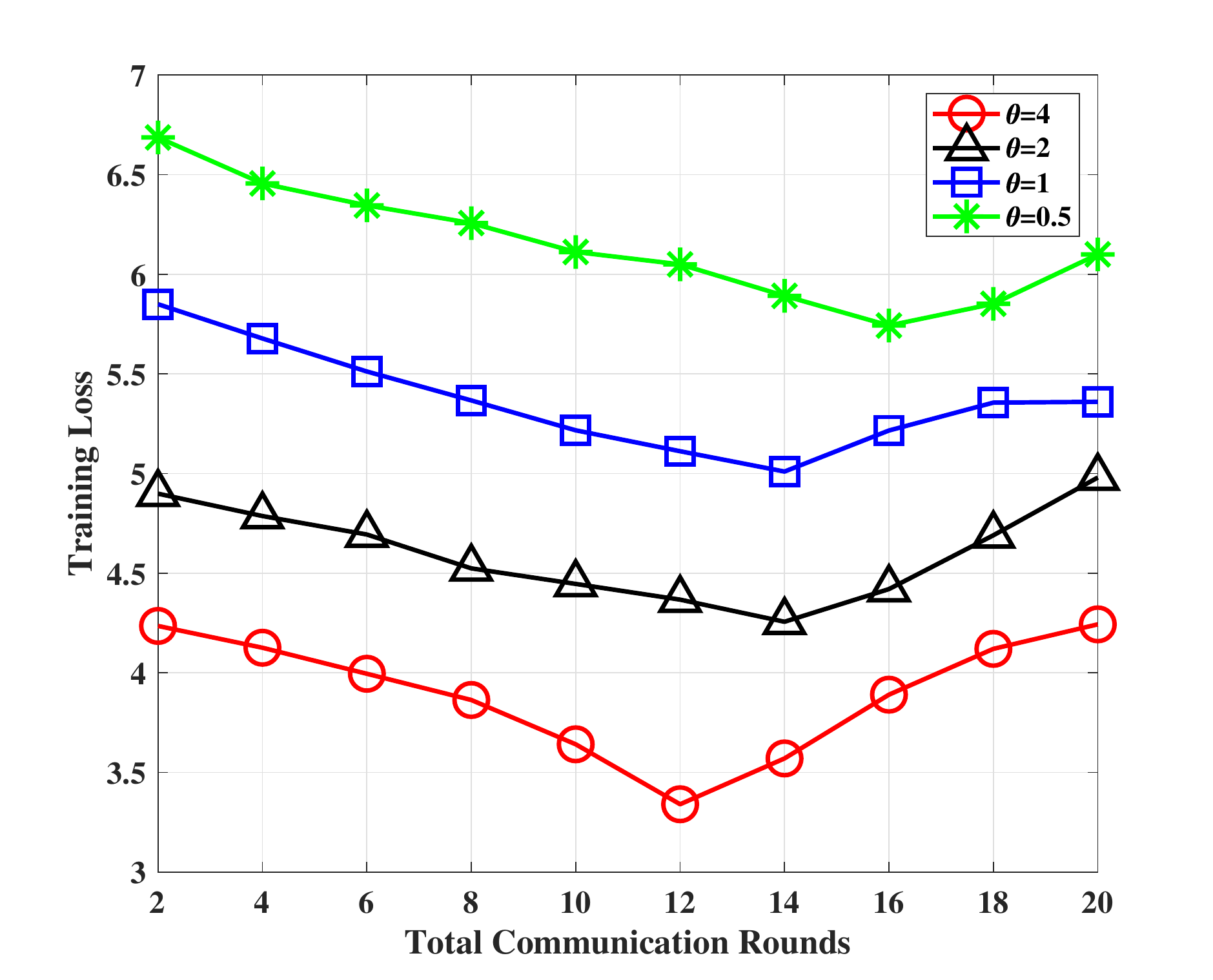}}
  \caption{Learning performance of different total communication rounds under different resource allocation ratios} \label{rs}
\end{figure}

In this subsection, we mainly present the results for the resource allocation, and the training loss value with different ratios ($\theta$) of both datasets are represented in the Fig.~\ref{rs}.

%\begin{figure}
%\centering
%  \includegraphics[width=0.4\textwidth]{try1.eps}
%  \caption{The system performances of different total communication rounds for different ratios} \label{try1}
%\end{figure}
As can be found in Fig.~\ref{rs}, the system performances for different ratios are investigated with the increasing number of total communication rounds. In details, we can find that there exists an optimal total communication round ($K$) for each computing ratio $\theta$. For example, the smallest training loss value can be obtained if clients end learning in 14 communication rounds with 15 learning epochs in each round when $\tau=1$ in the Fashion-MNIST dataset. Moreover, for different computing ratios, the optimal loss value tends to be different. This is due to the fact that the optimal number of local learning epoch is different with various $\theta$. In addition, similar trends can be found in the Cifar-10 dataset.
%although some highest performances are obtained in different total communication round. Mentioned here, this is a common situation in machine learning that the correlation between the loss value and testing accuracy may be not that consistent.

\subsection{Investigation on the lazy nodes}
\begin{figure}
\centering
\includegraphics[width=0.45\textwidth]{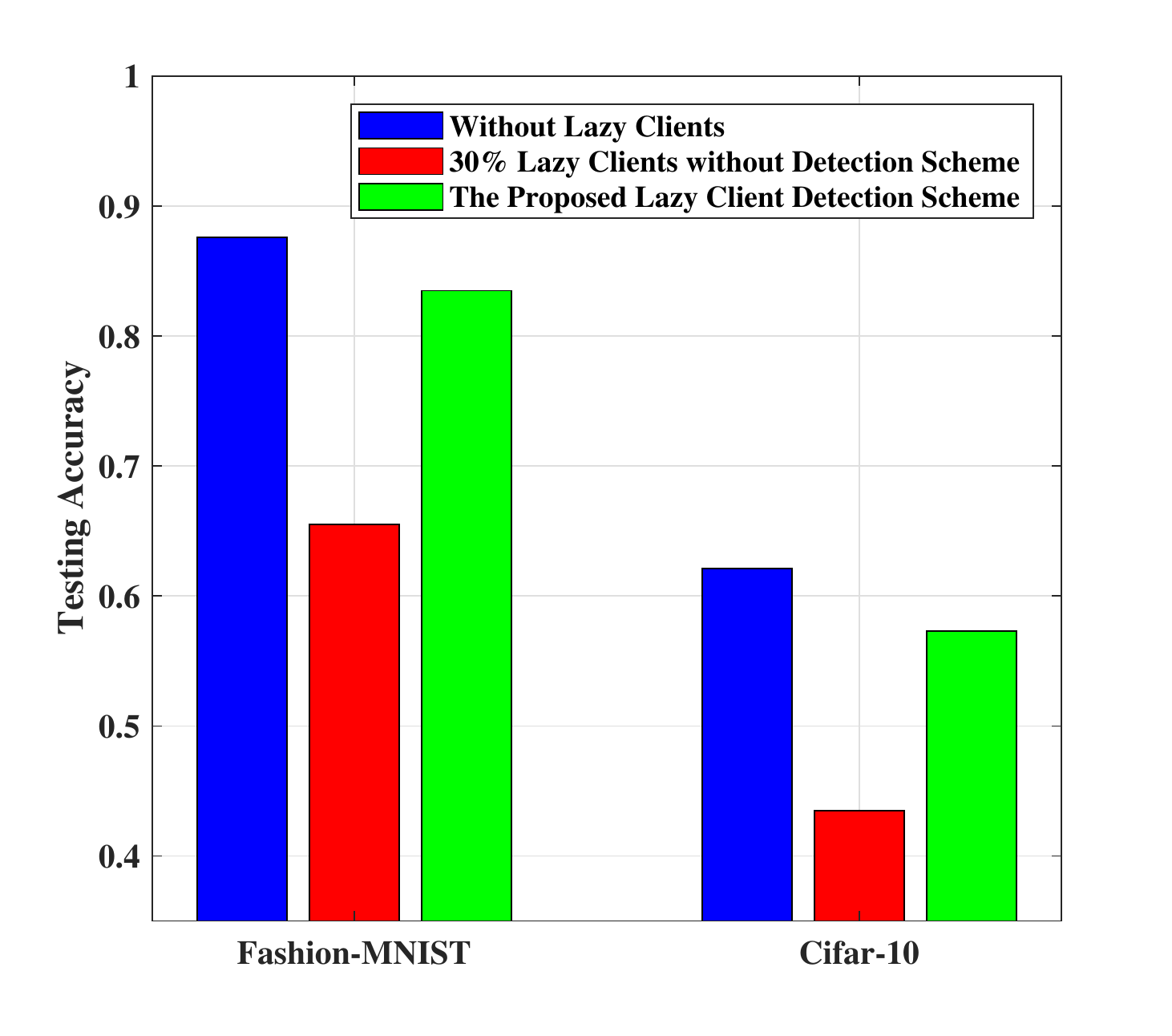}
\caption{Learning performance with/without lazy clients detection} \label{pnn}
\end{figure}

In this subsection, we investigate the impact of lazy clients on the proposed framework. We use signal to noise ratio (SNR) to denote the ratio between the power of original model parameters and that of the injected PN sequence, and Table~\ref{pn} represents the detecting rate of lazy clients under different SNRs. If the high peaks in terms of the cross-correlation coefficient surpass a predefined threshold, we can identify this client as a lazy one. We generate a $2^{15}$ length of PN sequence and use the first $25400$ values to add on the parameters. From the results with different SNRs, the detecting performances are remarkable and we can obtain a nearly $100\%$ rate to recognize the lazy clients when SNR=$3$ dB. Then Fig.~\ref{pnn} shows the PN sequence protecting performance (SNR=6 dB) when there are $30\%$ (6) lazy clients in each communication round. As can be found in this figure, the system performance with a certain percentage of lazy clients degrades sharply, i.e., $22.1\%$ and $19.6\%$ reduction in the Fashion-MNIST and Cifar-10 datasets, respectively. In addition, the proposed PN sequence protection method achieves $18\%$ and $13.8\%$ performance gain in each dataset, respectively.
\begin{table}
\centering
\caption{The detecting rate with different PN sequence power in the Fashion-mnist and Cifar-10 datasets}\label{pn}
\begin{tabular}{|c|c|c|c|}
  \hline
  {Signal to Noise Ratio} & 9 \textrm{dB} & 6 dB & 3 dB \\
  \hline
  Fashion-Mnist & 0.931 & 0.989 & 0.999 \\
  \hline
  Cifar-10 & 0.925 & 0.975 & 0.996\\
  \hline
\end{tabular}
\end{table}
%\begin{figure}
%\centering
%\subfigure[]{\label{lazy1}
%  \includegraphics[width=0.4\textwidth]{l1.pdf}}
%\subfigure[]{\label{lazy2}
%  \includegraphics[width=0.4\textwidth]{l2.pdf}}
%  \caption{Learning performance of different total communication rounds under different number of lazy clients} \label{lazy}
%\end{figure}
%\begin{figure}
%\centering
%  \includegraphics[width=0.4\textwidth]{try2.eps}
%  \caption{The system performances of different total communication rounds for different number of lazy nodes} \label{try2}
%\end{figure}
\section{Future Directions and Conclusion}
In this article, we have reviewed the weakness of FL and further investigated a blockchain assisted decentralized FL, called BLADE-FL. We then showed the effectiveness that the BLADE-FL can well address the potential issues, especially the single point of failure issue, existing in the traditional FL system. In addition, we have investigated the newly raising issues including privacy, resource allocation and lazy clients. Lastly, we have further provided related possible solutions and experimental results to solve these issues, which provide guidelines to the design of the BLADE-FL framework. For future directions, some asynchronous and heterogenetic investigations for different capabilities of clients, such as computing capability, training data size and transmitting diversity, as well as the smart contract design which provides reasonable rewards allocation for training and mining can be considered in the system. In addition, the light-weight model transmitting using quantization and sketch may be an alterative way to reduce the transmission cost.
\bibliographystyle{IEEEtran}
\bibliography{flb}

\begin{thebibliography}{10}
\providecommand{\url}[1]{#1}
\csname url@samestyle\endcsname
\providecommand{\newblock}{\relax}
\providecommand{\bibinfo}[2]{#2}
\providecommand{\BIBentrySTDinterwordspacing}{\spaceskip=0pt\relax}
\providecommand{\BIBentryALTinterwordstretchfactor}{4}
\providecommand{\BIBentryALTinterwordspacing}{\spaceskip=\fontdimen2\font plus
\BIBentryALTinterwordstretchfactor\fontdimen3\font minus
  \fontdimen4\font\relax}
\providecommand{\BIBforeignlanguage}[2]{{%
\expandafter\ifx\csname l@#1\endcsname\relax
\typeout{** WARNING: IEEEtran.bst: No hyphenation pattern has been}%
\typeout{** loaded for the language `#1'. Using the pattern for}%
\typeout{** the default language instead.}%
\else
\language=\csname l@#1\endcsname
\fi
#2}}
\providecommand{\BIBdecl}{\relax}
\BIBdecl

\bibitem{9048613}
C.~{Ma}, J.~{Li}, M.~{Ding}, H.~H. {Yang}, F.~{Shu}, T.~Q.~S. {Quek}, and H.~V.
  {Poor}, ``On safeguarding privacy and security in the framework of federated
  learning,'' \emph{IEEE Network}, vol.~34, no.~4, pp. 242--248, 2020.

\bibitem{8274963}
Z.~{Liu}, P.~{Longa}, G.~C. C.~F. {Pereira}, O.~{Reparaz}, and H.~{Seo},
  ``Four$\mathbb {Q}$q on embedded devices with strong countermeasures against
  side-channel attacks,'' \emph{IEEE Transactions on Dependable and Secure
  Computing}, vol.~17, no.~3, pp. 536--549, 2020.

\bibitem{9076082}
Y.~{Chen}, X.~{Qin}, J.~{Wang}, C.~{Yu}, and W.~{Gao}, ``Fedhealth: A federated
  transfer learning framework for wearable healthcare,'' \emph{IEEE Intelligent
  Systems}, vol.~35, no.~4, pp. 83--93, 2020.

\bibitem{9084352}
T.~{Li}, A.~K. {Sahu}, A.~{Talwalkar}, and V.~{Smith}, ``Federated learning:
  Challenges, methods, and future directions,'' \emph{IEEE Signal Processing
  Magazine}, vol.~37, no.~3, pp. 50--60, 2020.

\bibitem{8733825}
H.~{Kim}, J.~{Park}, M.~{Bennis}, and S.~{Kim}, ``Blockchained on-device
  federated learning,'' \emph{IEEE Communications Letters}, vol.~24, no.~6, pp.
  1279--1283, 2020.

\bibitem{8843900}
Y.~Lu, X.~Huang, Y.~Dai, S.~Maharjan, and Y.~Zhang, ``Blockchain and federated
  learning for privacy-preserved data sharing in industrial iot,'' \emph{IEEE
  Transactions on Industrial Informatics}, vol.~16, no.~6, pp. 4177--4186,
  2020.

\bibitem{8905038}
X.~{Bao}, C.~{Su}, Y.~{Xiong}, W.~{Huang}, and Y.~{Hu}, ``Flchain: A blockchain
  for auditable federated learning with trust and incentive,'' in \emph{2019
  5th International Conference on Big Data Computing and Communications
  (BIGCOM)}, 2019, pp. 151--159.

\bibitem{SHARMA2020102220}
P.~K. Sharma, J.~H. Park, and K.~Cho, ``Blockchain and federated learning-based
  distributed computing defence framework for sustainable society,''
  \emph{Sustainable Cities and Society}, vol.~59, p. 102220, 2020.

\bibitem{9051184}
S.~Wang, ``Blockfedml: Blockchained federated machine learning systems,'' in
  \emph{2019 International Conference on Intelligent Computing, Automation and
  Systems (ICICAS)}, 2019, pp. 751--756.

\bibitem{9134967}
Y.~Qu, S.~R. Pokhrel, S.~Garg, L.~Gao, and Y.~Xiang, ``A blockchained federated
  learning framework for cognitive computing in industry 4.0 networks,''
  \emph{IEEE Transactions on Industrial Informatics}, vol.~17, no.~4, pp.
  2964--2973, 2021.

\bibitem{inproceedings1}
S.~Otoum, I.~Al~Ridhawi, and H.~Mouftah, ``Blockchain-supported federated
  learning for trustworthy vehicular networks,'' 12 2020, pp. 1--6.

\bibitem{jelasity2011gossip}
M.~Jelasity, ``Gossip,'' in \emph{Self-organising Software}.\hskip 1em plus
  0.5em minus 0.4em\relax Springer, 2011, pp. 139--162.

\bibitem{8869754}
L.~{Ismail}, H.~{Materwala}, and S.~{Zeadally}, ``Lightweight blockchain for
  healthcare,'' \emph{IEEE Access}, vol.~7, pp. 149\,935--149\,951, 2019.

\bibitem{8048837}
P.~{Fairley}, ``Blockchain world - feeding the blockchain beast if bitcoin ever
  does go mainstream, the electricity needed to sustain it will be enormous,''
  \emph{IEEE Spectrum}, vol.~54, no.~10, pp. 36--59, 2017.

\bibitem{9347706}
K.~Wei, J.~Li, M.~Ding, C.~Ma, H.~Su, B.~Zhang, and H.~V. Poor, ``User-level
  privacy-preserving federated learning: Analysis and performance
  optimization,'' \emph{IEEE Transactions on Mobile Computing}, pp. 1--1, 2021.

\end{thebibliography}
\end{document}